%% file: main_sigir.tex
\documentclass[sigconf]{acmart}
\usepackage{color}
\usepackage{multirow}
\usepackage{subfigure}
\usepackage{enumitem}

\newcommand{\modelname}{JTCN}

\AtBeginDocument{%
  \providecommand\BibTeX{{%
    \normalfont B\kern-0.5em{\scshape i\kern-0.25em b}\kern-0.8em\TeX}}}

\setcopyright{acmcopyright}
\copyrightyear{2020}
\acmYear{2020}
\acmDOI{10.1145/XXXXXX.XXXXXX}

\acmConference[SIGIR'20]{43rd International ACM SIGIR Conference on Research and Development in Information Retrieval}{July 25--30, 2020}{Xi'an, China}
\acmBooktitle{43rd International ACM SIGIR Conference on Research and Development in Information Retrieval (SIGIR '20), July 25--30, 2020, Xi'an, China}
\acmPrice{15.00}
\acmISBN{978-1-4503-XXXX-X/18/06}
\begin{document}
\fancyhead{}
%%
%% The "title" command has an optional parameter,
%% allowing the author to define a "short title" to be used in page headers.
% \title{Cold Start Recommendation with Zero-shot Capsule}
%\title{A Joint Training Cold-start Recommendation Framework Network with Capsule Network.}
\title{Joint Training Capsule Network for Cold Start Recommendation}

%%
%% The "author" command and its associated commands are used to define
%% the authors and their affiliations.
%% Of note is the shared affiliation of the first two authors, and the
%% "authornote" and "authornotemark" commands
%% used to denote shared contribution to the research.
%\author{Anonymous Author(s)}、、
\settopmatter{authorsperrow=4}

\author{Tingting Liang}
%\authornote{Both authors contributed equally to this research.}
\affiliation{%
  \institution{Hangzhou Dianzi University}
  \city{Hangzhou}
  \country{China}
}
\email{liangtt@hdu.edu.cn}

\author{Congying Xia}
\affiliation{%
  \institution{University of Illinois at Chicago}
  \city{Chicago}
  \country{US}}
\email{cxia8@uic.edu}

\author{Yuyu Yin}
%\authornote{Both authors contributed equally to this research.}
\affiliation{%
  \institution{Hangzhou Dianzi University}
  \city{Hangzhou}
  \country{China}
}
\email{yinyuyu@hdu.edu.cn}

\author{Philip S. Yu}
\affiliation{%
  \institution{University of Illinois at Chicago}
  \city{Chicago}
  \country{US}}
\email{psyu@uic.edu}

%%
%% By default, the full list of authors will be used in the page
%% headers. Often, this list is too long, and will overlap
%% other information printed in the page headers. This command allows
%% the author to define a more concise list
%% of authors' names for this purpose.
% \renewcommand{\shortauthors}{Trovato and Tobin, et al.}

%%
%% The abstract is a short summary of the work to be presented in the
%% article.
\begin{abstract}
This paper proposes a novel neural network, joint training capsule network (\modelname), for the cold start recommendation task.
%We propose to use an attentive capsule layer to extract multiple user preferences from the interaction history via a dynamic routing-by-agreement mechanism. 
We propose to mimic the high-level user preference other than the raw interaction history based on the side information for the fresh users.
Specifically, an attentive capsule layer is proposed to aggregate high-level user preference from the low-level interaction history via a dynamic routing-by-agreement mechanism.
Moreover, {\modelname} jointly trains the loss for mimicking the user preference and the softmax loss for the recommendation together in an end-to-end manner.
Experiments on two publicly available datasets demonstrate the effectiveness of the proposed model. {\modelname} improves other state-of-the-art methods at least 7.07\% for CiteULike and 16.85\% for Amazon in terms of Recall@100 in cold start recommendation.
\end{abstract}

%%
%% The code below is generated by the tool at http://dl.acm.org/ccs.cfm.
%% Please copy and paste the code instead of the example below.
%%
\begin{CCSXML}
<ccs2012>
<concept>
<concept_id>10002951.10003317.10003347.10003350</concept_id>
<concept_desc>Information systems~Recommender systems</concept_desc>
<concept_significance>500</concept_significance>
</concept>
<concept>
<concept_id>10010147.10010257.10010293.10010294</concept_id>
<concept_desc>Computing methodologies~Neural networks</concept_desc>
<concept_significance>500</concept_significance>
</concept>
</ccs2012>
\end{CCSXML}

\ccsdesc[500]{Information systems~Recommender systems}
\ccsdesc[500]{Computing methodologies~Neural networks}

%%
%% Keywords. The author(s) should pick words that accurately describe
%% the work being presented. Separate the keywords with commas.
\keywords{Recommender systems, Cold start, User preference estimation}

%% A "teaser" image appears between the author and affiliation
%% information and the body of the document, and typically spans the
%% page.
% \begin{teaserfigure}
%   \includegraphics[width=\textwidth]{sampleteaser}
%   \caption{Seattle Mariners at Spring Training, 2010.}
%   \Description{Enjoying the baseball game from the third-base
%   seats. Ichiro Suzuki preparing to bat.}
%   \label{fig:teaser}
% \end{teaserfigure}

%%
%% This command processes the author and affiliation and title
%% information and builds the first part of the formatted document.
\maketitle

\input{1intro.tex}

\input{2model.tex}
\input{3exp.tex}
\input{4con.tex}

%%
%% The acknowledgments section is defined using the "acks" environment
%% (and NOT an unnumbered section). This ensures the proper
%% identification of the section in the article metadata, and the
%% consistent spelling of the heading.
%\vspace{-0.1in}
\begin{acks}
% We thank the reviewers for their valuable comments.
This work is supported in part by NSFC under grant 61872119, NSF under grants III-1526499, III-1763325, III-1909323, CNS-1930941, and Key Research \& Development Plan of Zhejiang Province under grant 2019C03134.  

\end{acks}

%%
%% The next two lines define the bibliography style to be used, and
%% the bibliography file.
\bibliographystyle{ACM-Reference-Format}
\bibliography{acmart_abbr}
%%
%% If your work has an appendix, this is the place to put it.
\appendix

\end{document}

%% file: 1intro.tex
\section{Introduction}
The effectiveness of current recommender systems highly relies on the interactions between users and items. These systems usually do not perform well when new users or new items arrive. This challenge is widely known as cold start recommendation \cite{lin2013addressing}. To alleviate this problem, models have been proposed to leverage side information such as user attributes \cite{fernandez2016alleviating} or user social network data \cite{lin2013addressing,sedhain2017low} to generate recommendations for new users. These models can be grouped into three categories based on how they use the side information: similarity-based models \cite{sarwar2001item} which calculate similarities between items based on the side information; matrix factorization methods with regularization \cite{he2016vbpr} that regularize the latent features based on auxiliary relations; matrix factorization methods with feature mapping \cite{gantner2010learning} which learn a mapping between the side information and latent features. According to \cite{li2019zero}, these cold start models can be viewed within a simple unified linear framework which learns a mapping between the side information and the interaction history.

Recently, deep learning models (DNNs) have emerged to tackle the cold start problem by providing a larger model capacity.
Volkovs et al. \cite{volkovs2017dropoutnet} regard cold start recommendation as a data missing problem and modifies the learning procedure by applying dropout to input mini-batches. It highly depends on the generalization ability of the dropout technique to generalize the model from warm start to cold start.
\cite{li2019zero} is the first work that proposes to solve the cold start problem in the Zero-shot Learning (ZSL) perspective. It leverages a low-rank auto-encoder to reconstruct interaction history from the user attributes. However, it is a two-step method which firstly learns the reconstruction and then solves the recommendation for the cold start users or items in the second step. A two-step method might suffer from the error propagation problem.

To avoid the aforementioned problems and fully understand the content in the side information, we propose an end-to-end joint training capsule network (\modelname) for cold start recommendation. In {\modelname}, a user is represented explicitly in two folds: the high-level user preference and the content contained in the side information. The high-level user preference is aggregated from the low-level interaction history through the attentive capsule layer with a dynamic routing-by-agreement mechanism \cite{sabour2017dynamic, xia2018zero}.

A mimic loss is proposed to mimic the high-level user preference for cold start users or items from the side information. We argue that it is more explainable to mimic the high-level user preference than the low-level interaction history. It would be natural to infer user preference from the side information other than non-existent interaction history.
Another softmax loss is used to train the regular recommendation process. Our goal is to not only mimic the high-level user preference for the cold start users or items, but also effectively do recommendations for them. We propose to achieve our goals by jointly training these two losses together in an end-to-end manner. %Experiments on two real-world datasets show that our proposed model outperforms baselines consistently for the cold start recommendation task. 
\begin{figure*}[ht]
    \centering
    \includegraphics[width=13.8cm]{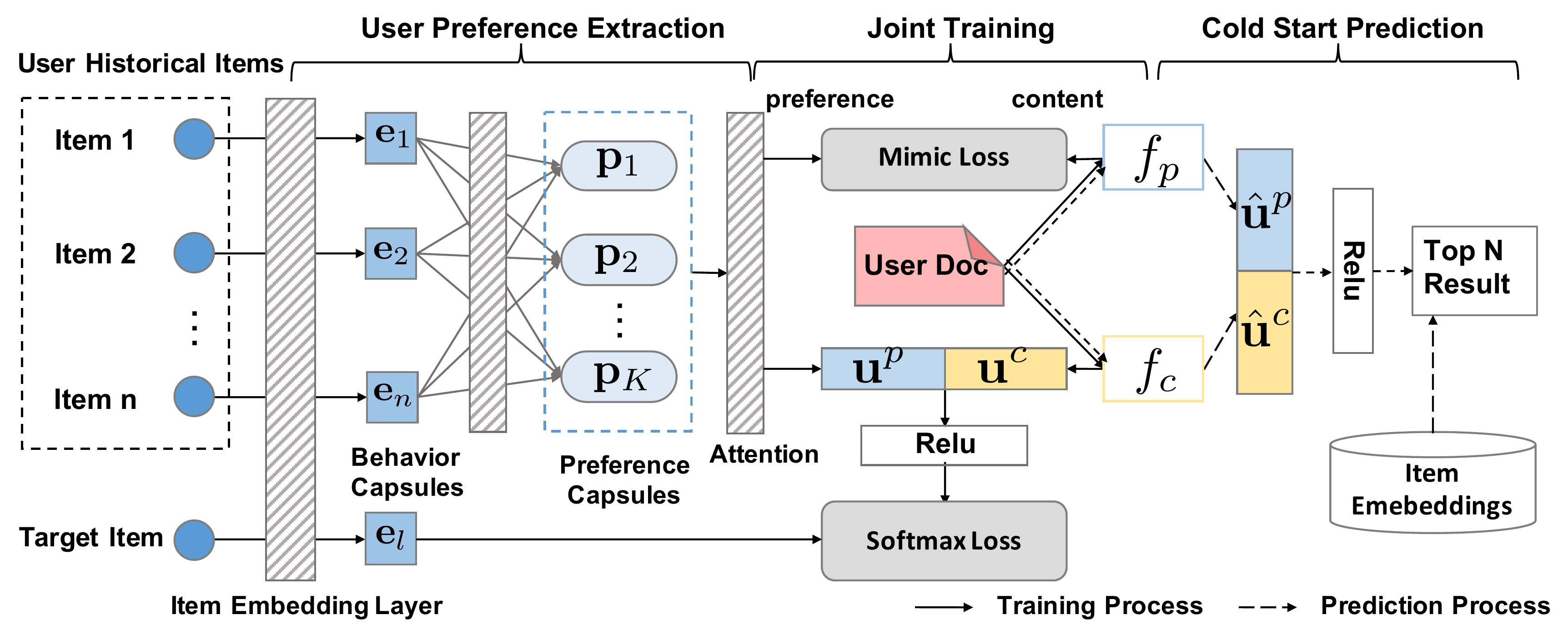}
    \vskip -0.12in
    \caption{The architecture of {\modelname}. 
    % {\color{red}xxx} takes user behaviors and profile features as inputs to respectively output user preference and content representations, which can be used for retrieving items for cold start users.
    Id features of items are transformed into embeddings through the embedding layer. The embeddings of historical items are fed into the attentive multi-preference extraction layer, which consists of one multi-preference extraction layer and one attention layer, to obtain the user preference representation. The content representation produced by network $f_{c}$ is fused with the preference embedding to form the softmax loss. The output of network $f_{p}$ is used to approximate user preference through a mimic loss. When making predictions for a new user, $f_{c}$ and $f_{p}$ is able to generate a comprehensive representation for the user based on the side information, including both preference and content information. }
        \vskip -0.12in
    \label{fig:model}
    
\end{figure*}
In summary, the contributions of this paper are:
\begin{itemize}[leftmargin=*]
    \item \textbf{Joint Training}: A joint training framework is proposed for the cold start recommendation by training the mimic loss for the cold start and the softmax loss for the recommendation together.
    \item \textbf{Capsule Network}: An attentive capsule layer is proposed to aggregate high-level user preference from the low-level interaction history via a dynamic routing-by-agreement mechanism.
    \item \textbf{Demonstrated Effectiveness}: Experiments on two real-world datasets show that our proposed model outperforms baselines consistently for the cold start recommendation task.
\end{itemize}

%% file: 2model.tex
% \subsection{Problem Formalization}

\vspace{-0.15in}
\section{Proposed Model}
\vspace{-0.05in}
\subsection{Problem Statement}
\vspace{-0.05in}
We consider a recommender system with a user set $\mathcal{U}=\{u_1,\dots,u_N\}$ and an item set $\mathcal{V}=\{v_1,\dots,v_M\}$, where N is the number of users and M is the number of items. The user-item feedback can be represented by a matrix $R\in\mathbb{R}^{N\times M}$, where $R_{ij}=1$ if user $i$ gives a positive feedback on item $j$, and $R_{ij}=0$ otherwise. Let $\mathcal{U}(j)=\{i\in\mathcal{U}|R_{ij}\neq0\}$ be the set of users that had shown preference to item $j$, and $\mathcal{V}(i)=\{j\in\mathcal{V}|R_{ij}\neq0\}$ be the set of items that user $i$ gave positive feedback. This paper focuses on the cold start scenario in which no preference clue is available, namely, $\mathcal{V}(i)=\varnothing$ or $\mathcal{U}(j)=\varnothing$ for a given user $i$ or given item $j$. Our objective is to generate personalized recommendation results for each fresh user or item based on its corresponding side information.
% We aim to provide accurate recommendation results for a fresh user or item based on its corresponding side information. 
% Formally, the problem can be defined 
% \subsection{The Architecture of XXX}

\vspace{-0.1in}
\subsection{User Multi-Preference Extraction}
\vspace{-0.05in}
The framework of our proposed model is illustrated in Figure \ref{fig:model}. The framework here is mainly for cold start users, and the framework for cold start items can be modeled in the same manner. 
As shown in Figure \ref{fig:model}, the input of {\modelname} consists of user documents which contain the side information (labeled as User Doc in the figure), historical items, and the target item. 
The former two can be respectively used for extracting user properties and preferences. 
The target item is the one that we use to make a prediction for the user during the training process.
The usage of user documents will be discussed in Section \ref{subsec:training}. This section focuses on the extraction of high-level user preferences.
\vspace{-0.1in}
\subsubsection{Embedding Layer}
The user preference extraction part starts with the item embedding layer which embeds the id features of items into low-dimensional dense vectors. For the target item, the embedding is denoted as $\mathbf{e}_l\in\mathbb{R}^{d}$. For the historical items of user $i$ (\emph{i.e.,} $\mathcal{V}(i)$), corresponding item embeddings are gathered to form the set of user preference embeddings $\mathrm{E}_i=\{\mathbf{e}_j, j\in\mathcal{V}(i)\}$. 
% The embedding of the label item is denoted as $\mathbf{e}_i$.
\vspace{-0.1in}
\subsubsection{Attentive Capsule Layer for Multi-preference Extraction}
An important task of our {\modelname} is to learn a network for mimicking high-level preference representations for cold start users from their side information. Therefore, it is crucial to construct a representative user preference embedding during the training process. Representing user preference by a simple combination (\emph{e.g.,} averaging, concatenation) of vectors $\mathbf{e}_j\in\mathrm{E}_i$ is not conducive to extracting diverse interests of users. Inspired by \cite{li2019multi}, we propose to apply the recently proposed dynamic routing in capsule network \cite{sabour2017dynamic} to capture multiple preferences for users. 
Considering that not all the preference capsules contribute equally to aggregate the high-level user preference representation. We further propose to adopt the attention mechanism to discriminate the informative capsules.
% {\color{red}[connection between dynamic routing and multi-preference embedding]}

We consider two layers of capsules, which we name as behavior capsules and preference capsules, to represent the user behavior (historical items) and multiple user preferences respectively. Dynamic routing is adopted to compute the vectors of preference capsules based on the vectors of behavior capsules in an iterative way. In each step, given the embedding $\mathbf{e}_{j}\in\mathbb{R}^{d}$ of behavior capsule $j$ and vector $\mathbf{p}_{k}\in\mathbb{R}^{d}$ of preference capsule $k$, the routing logit is calculated by
\vspace{-0.05in}
\begin{equation}
    b_{jk} = \mathbf{p}_{k}^{T}\mathbf{S} \mathbf{e}_{j},
    \vspace{-0.03in}
\end{equation}
where $\mathbf{S}\in\mathbb{R}^{d\times d}$ denotes the bilinear mapping matrix parameter shared across each pair of behavior and preference capsules.

The coupling coefficients between behavior capsule $j$ and all the preference capsules sum to 1 and are determined by performing the ``routing softmax'' on logits as:
\vspace{-0.05in}
\begin{equation}
    c_{jk} = \frac{\mathrm{exp}(b_{jk})}{\sum_{t}{\mathrm{exp}(b_{jt})}}.
    \vspace{-0.03in}
\end{equation}
With the coupling coefficients calculated, the candidate vector for preference capsule $k$ is computed by the weighted sum of all behavior capsules:
\vspace{-0.05in}
\begin{equation}
    \mathbf{z}_k = \sum_{j}{c_{jk}\mathbf{S}\mathbf{e}_j}.
    \vspace{-0.05in}
\end{equation}
The embedding of preference capsule $k$ is obtained by a non-linear ``squash'' function as:
\vspace{-0.08in}
\begin{equation}
    \mathbf{p}_{k} = squash(\mathbf{z}_k) = \frac{||\mathbf{z}_k||^2}{1+||\mathbf{z}_k||^2}\frac{\mathbf{z}_k}{||\mathbf{z}_k||}.
\end{equation}
\vspace{-0.05in}

Suppose we have $K$ preference capsules, which means there are $K$ distinct preferences of users extracted from the historical items on which the users gave positive feedback. We apply an attention layer to emphasize the informative capsules. There exist several effective ways to calculate the attention score and this paper adopts the multi-layer perceptron (MLP) as
\vspace{-0.05in}
\begin{equation}
    a_{k} = \mathbf{h}^{T}\mathrm{ReLU}(\mathbf{W}_a\mathbf{p}_{k}+\mathbf{b}_a),
    \hspace{1em}
    \alpha_{k} = \frac{\mathrm{exp}(a_k)}{\sum_{k=1}^{K}{\mathrm{exp}(a_k)}},
    \vspace{-0.03in}
\end{equation}
% \vspace{-0.03in}
where $\mathbf{W}_a\in\mathbb{R}^{d\times d_{a}}$, $\mathbf{b}_a\in\mathbb{R}^{d_a}$, and $\mathbf{h}\in\mathbb{R}^{d_a}$ are the attention layer parameters. The final attentive weight is normalized by the softmax function.
% \begin{equation}
%     \alpha_{k} = \frac{\mathrm{exp}(a_k)}{\sum_{k=1}^{K}{\mathrm{exp}(a_k)}}.
% \end{equation}

With the attentive weights assigned to the preference capsules, the high-level user preference can be formed as the weighted sum: 
\vspace{-0.03in}
\begin{equation}\label{eq:pre_emb}
    \mathbf{u}^{p} = \sum_{k=1}^{K}{\alpha_k\mathbf{p}_k}.
\end{equation}
\vspace{-0.08in}

\vspace{-0.15in}
\subsection{Joint Training}\label{subsec:training}
\vspace{-0.05in}
%In the cold start scenario, only the side information of users is available. In addition to the user preference information, the content information extracted from user profiles plays a great role in recommender system, especially in the cold star scenarios. 
%Also, two neural networks that are capable of mapping the user profile into one content space and one preference space, in which the corresponding representations could be used for the final prediction.
%Our goal is to design a joint loss to learn two networks 
A joint training framework is proposed here by optimizing two losses together: a softmax loss for recommendation and a mimic loss for generating user preferences for cold start users without interaction history. The user representation in {\modelname} is represented in two folds, the user preferences and the content. Two MLP networks, namely $f_{c}$ and $f_{p}$, are used to map the user document into one content space and one preference space for the cold start users. Those two representations are fused together for the 
%, in which the corresponding representations could be used for 
the final prediction. 
%Given the user profile, we propose to pass it through two MLP networks, 
\subsubsection{Softmax Loss}
The output of $f_{c}$ denoted by $\mathbf{u}^{c}$ is fused together with the high-level user preference embedding defined by (\ref{eq:pre_emb}) to form the user embedding. The representation of user $i$ can be generated by
\vspace{-0.1in}
\begin{equation}
    \mathbf{u}_i = \mathrm{ReLU}(\mathbf{W}_{u}[f_{c}(\mathrm{X}_{i}), \mathbf{u}_{i}^{p}]+\mathbf{b}_{u}),
    \vspace{-0.03in}
\end{equation}
% \vspace{-0.05in}
where $\mathrm{X}_{i}$ denotes the input user document, $\mathbf{W}_{u}\in\mathbb{R}^{d\times 2d}$ and $\mathbf{b}_{u}\in\mathbb{R}^{d}$ are the parameters.
$[\cdot,\cdot]$ denotes the concatenation operation and $\mathrm{ReLU}(\cdot)$ is the Rectified Linear Unit.
With the user vector $\mathbf{u}_i$ and the target item embedding $\mathbf{e}_{l}$, the probability of the user interacting with the target item can be predicted by
\vspace{-0.05in}
\begin{equation}
\mathrm{Pr}(\mathbf{e}_{l}|\mathbf{u}_i)=\frac{\mathrm{exp}(\mathbf{u}_{i}^{\mathrm{T}}\mathbf{e}_{l})}{\sum_{j\in\mathcal{V}(i)}\mathrm{exp}(\mathbf{u}_{i}^{\mathrm{T}}\mathbf{e}_{j})}.
\vspace{-0.03in}
\end{equation}
We use the \textit{softmax loss} as the objective function to minimize for the recommendation training:
\vspace{-0.05in}
\begin{equation}
    \mathcal{L}_{softmax}=-\sum_{(i,l)\in\mathcal{D}}\mathrm{log}\mathrm{Pr}(\mathbf{e}_{l}|\mathbf{u}_i),
    \vspace{-0.03in}
\end{equation}
where $\mathcal{D}$ is the collection of training data containing user-item interactions.
% \vspace{-0.05in}
\subsubsection{Mimic Loss}
\vspace{-0.05in}
In order to learn preference information from the document of a new user, we propose to use the output of $f_{p}$ to approximate the high-level user preference representation defined by (\ref{eq:pre_emb}).
We define the \textit{mimic loss} as the mean square difference as:
\vspace{-0.08in}
\begin{equation}
    \mathcal{L}_{mimic} =\frac{1}{|\mathcal{D}|}\sum_{(i,l)\in\mathcal{D}}\sum_{d} (\mathbf{u}^{p}_{i}-f_{p}(\mathrm{X}_{i}))^2.
\end{equation}
\vspace{-0.08in}

% Combining the softmax and mimic loss into a joint loss as follows 
Jointly training the following combination of softmax loss and mimic loss enables the network to better imitate the high-level preference and capture content from the side information of new users, which greatly improve the cold start recommendation performance:
\vspace{-0.08in}
\begin{equation}
    \mathcal{L}_{joint} = \mathcal{L}_{softmax}+\mathcal{L}_{mimic}.
\end{equation}
\vspace{-0.1in}

\vspace{-0.08in}
\subsection{Cold Start Prediction}
\vspace{-0.03in}
Once training is completed, as shown in the right side of Figure \ref{fig:model}, we fix the model and make a forward pass through $f_{c}$ and $f_{p}$ to get the representation for a new user based on its side information as:
% \begin{equation}
%     \mathbf{u}_{new} = \mathrm{ReLU}(\mathbf{W}_{u}[f_{c}(\mathrm{X}_{new}),f_{p}(\mathrm{X}_{new})]+\mathbf{b}_{u}).
% \end{equation}
\vspace{-0.05in}
\begin{equation}
    \mathbf{u}_{new} = \mathrm{ReLU}(\mathbf{W}_{u}[\hat{\mathbf{u}}^{c},\hat{\mathbf{u}}^{p}]+\mathbf{b}_{u}),
\end{equation}
where $\hat{\mathbf{u}}^{c}=f_{c}(\mathrm{X}_{new})$ and $\hat{\mathbf{u}}^{p}=f_{p}(\mathrm{X}_{new})$.
At last, the preference score of the new user on item $j$ is decided by the inner product of the corresponding embeddings:
\vspace{-0.05in}
\begin{equation}
    \hat{r}_{new,j} = \mathbf{u}_{new}^{\mathrm{T}}\mathbf{e}_{j}.
\end{equation}
\vspace{-0.05in}

%% file: 3exp.tex
\vspace{-0.15in}
\section{Experiments}
\vspace{-0.05in}

% \subsection{Dataset and Experimental Setup}
\subsection{Datasets}
\vspace{-0.05in}
% We choose two publicly available datasets: CiteULike\footnote{http://www.citeulike.org.} and Amazon Movies and TV\footnote{http://jmcauley.ucsd.edu/data/amazon/} \cite{ni2019justifying}, for evaluating cold start recommendation performance. CiteULike contains implicit user-article feedback, Both of the datasets contain item content formation in the form of title and abstract/description. 
% For CiteULike, we use a vocabulary of top 8,000 words selected by tf-idf \cite{wang2011collaborative}. The vocabulary size of Amazon is 10,000. 

We choose two public datasets for evaluating cold start recommendation performance. 1) CiteULike\footnote{http://www.citeulike.org.} with 5,551 users, 16,980 articles, and 204,986 implicit user-article feedbacks. CiteULike contains article content formation in the form of title and abstract. We use a vocabulary of the top 8,000 words selected by tf-idf \cite{wang2011collaborative}. 2) Amazon Movies and TV\footnote{http://jmcauley.ucsd.edu/data/amazon/} \cite{ni2019justifying}. We convert the explicit feedbacks with rating 5 to implicit feedbacks.
We filter the user and items with interactions less than 10 and finally get 14,850 users, 23,232 items, and 548,296 interactions. The vocabulary size of words selected by tf-idf from item titles and descriptions is 10,000.

Since only item side information is available, we recommend users for the cold start items. For both datasets, we randomly select 20\% of items as the cold start items which will be recommended users at test time. We use Recall and NDCG as evaluation metrics.

\vspace{-0.1in}
\subsection{Baselines}
\vspace{-0.05in}
We compare the proposed {\modelname} with several representative recommendation models including three content-based methods \textbf{KNN}~\cite{sarwar2001item}, \textbf{FM}~\cite{rendle2010factorization}, and \textbf{VBPR}~\cite{he2016vbpr}, two deep learning methods \textbf{LLAE}~\cite{li2019zero} and \textbf{DropoutNet}~\cite{volkovs2017dropoutnet}. 
KNN uses content information to compute the cosine similarity between items. DropoutNet uses WMF~\cite{hu2008collaborative} as the pre-trained model for input preference.
Except for KNN and LLAE which do not have the parameter of latent factor, we set the number of latent factors $d=256$ for all methods. The other hyperparameters of all the compared methods are tuned to find an optimal result. 
For {\modelname}, the number of preference capsules is set $K=5$, the dimension of attention layer is set $d
_a=128$, and the \emph{Adam} optimizer with the learning rate of 0.0005 is adopted.
For all the methods except KNN, we use the early stopping strategy with a patience of 10.
% \textbf{FM} \cite{rendle2010factorization} is a widely used
% content-aware recommendation method which explores pairwise
% interactions between item features. \textbf{VBPR} \cite{he2016vbpr} is a visual personalized ranking model that uncovers visual and latent dimensions simultaneously.
% \textbf{LLAE} From Zero-Shot Learning to Cold-Start Recommendation \citet{li2019zero}
% \textbf{DropoutNet}: \cite{volkovs2017dropoutnet} 
% \textbf{KNN} \cite{sarwar2001item} is a neighborhood-based method using collaborative
% user similarities.

% {\color{red}Implementation Details}

\vspace{-0.1in}
\subsection{Results}
% \vspace{-0.01in}
% Table generated by Excel2LaTeX from sheet 'citeu-final'

% \vspace{-0.1in}
\begin{table}[t]
  \centering
  \caption{Performance Comparison on two datasets in terms of Recall@100 and NDCG@100.}
    \vspace{-0.15in}
    \begin{tabular}{l|cc|cc}
    \toprule
    \multirow{2}[3]{*}{Methods} & \multicolumn{2}{c|}{CiteULike} & \multicolumn{2}{c}{Amazon } \\
\cmidrule{2-5}          & Recall & NDCG & Recall & NDCG \\
\midrule
    KNN   & 0.2981 & 0.3453 & 0.0564 & 0.2358 \\
    FM    & 0.5100 & 0.4583 & 0.0924 & 0.2260 \\
    VBPR  & 0.5426 & 0.4825 & 0.0891 & 0.2215 \\
    LLAE  & 0.5816 & 0.5286* & 0.1264* & 0.2439 \\
    DropoutNet & 0.6011* & 0.5226 & 0.1013 & 0.2815*  \\
    \midrule
    JTCN  & \textbf{0.6436} & \textbf{0.5432} & \textbf{0.1477} & \textbf{0.3364} \\
    Improve & 7.07\% & 2.76\%& 16.85\%& 19.50\%\\
    \bottomrule
    \end{tabular}%
    %\vspace{-0.02in}
  \label{tab:performance}%
  \vskip -0.2in
\end{table}%

\subsubsection{Model Comparison}
The experimental results of our {\modelname} as well as baselines on two datasets are reported in Table \ref{tab:performance} in terms of Recall@100 and NDCG@100 (with $d=256$). The best results are listed in bold, and the second best results are marked with star (*). Clearly, {\modelname} remarkably outperforms baseline models on both datasets. KNN shows poor performance in the cold start scenario, which led by the rough estimation of content-based similarity without any historical interaction. 
% FM and VBPR give comparable performance, especially for Amazon dataset. 
The improvement obtained by FM compared with KNN indicates the advantage of feature interaction. VBPR performs slightly better as it is proposed to alleviate the cold start problem by using both latent factors and content factors that are extracted from auxiliary information \cite{he2016vbpr}. It can be easily observed that deep learning based methods, LLAE and DropoutNet, which are dedicated to cold start problem, perform better than the traditional content-based baselines.
However, all the baselines only reconstruct or extract content factors from the input side information in the test stage.
% even though some of them consider both content and preference information during training.
{\modelname} outperforms all baselines improving Recall@100 by 7.07\% and 16.85\% on two datasets over the best baseline. This indicates that combining the content information with preference information generated based on the raw input of new users or items can effectively improve the performance of cold start 
recommendation. 

In addition, DropoutNet has a need for the pre-trained model to generate preference input for the main DNN, which may limit its generalization on different datasets. 
In contrast, the proposed {\modelname} doesn't need such a pre-trained model to handle the input by learning directly from the input raw features. 

% % Table generated by Excel2LaTeX from sheet 'citeu-final'
% \begin{table}[htbp]
%   \centering
%   \caption{Cold-start recommendation results on CiteULike}
%     \begin{tabular}{l|cc|cc}
%     \toprule
%     Methods & R@50  & R@100 & N@50  & N@100 \\
%     \midrule
%     KNN   & 0.2118 & 0.2981 & 0.3348 & 0.3453 \\
%     FM    & 0.3743 & 0.5100 & 0.4307 & 0.4583 \\
%     VBPR  & 0.4058 & 0.5426 & 0.4863 & 0.4825 \\
%     LLAE  & 0.4639 & 0.5816 & 0.5333 & 0.5286 \\
%     DropoutNet & 0.4639 & 0.6011 & 0.5278 & 0.5226 \\
%     \midrule
%     {\color{red}Ours}  & \textbf{0.5031} & \textbf{0.6436} & \textbf{0.5495} & \textbf{0.5432} \\
%     % bert  & \textbf{0.5179} & \textbf{0.6588} & \textbf{0.5584} & \textbf{0.5513} \\
%     \bottomrule
%     \end{tabular}%
%   \label{tab:addlabel}%
% \end{table}%

% % Table generated by Excel2LaTeX from sheet 'amazon'
% \begin{table}[htbp]
%   \centering
%   \caption{Add caption}
%     \begin{tabular}{l|cc|cc}
%     \toprule
%     Methods & R@50  & R@100 & N@50  & N@100 \\
%     \midrule
%     KNN   & 0.0393 & 0.0564 & 0.2247 & 0.2358 \\
%     FM    & 0.0629 & 0.0924 & 0.2121 & 0.2260 \\
%     VBPR  & 0.0599 & 0.0891 & 0.2086 & 0.2215 \\
%     LLAE  & 0.0908 & 0.1264 & 0.2294 & 0.2439 \\
%     DropoutNet &       &       &       &  \\
%     \midrule
%     Ours  & \textbf{0.1026} & \textbf{0.1477} & \textbf{0.3286} & \textbf{0.3364} \\
%     \bottomrule
%     \end{tabular}%
%   \label{tab:addlabel}%
% \end{table}%

\vspace{-0.05in}
\subsubsection{Impact of Latent Factors}
\begin{figure}[ht]
\vspace{-0.1in}
\centering
% \subfigure{\includegraphics[width=4.3cm]{icml2019/fig/dynamic_fig/item_gray_3_50.pdf}}
\subfigure[CiteULike--Recall@100]{\includegraphics[width=4.15cm]{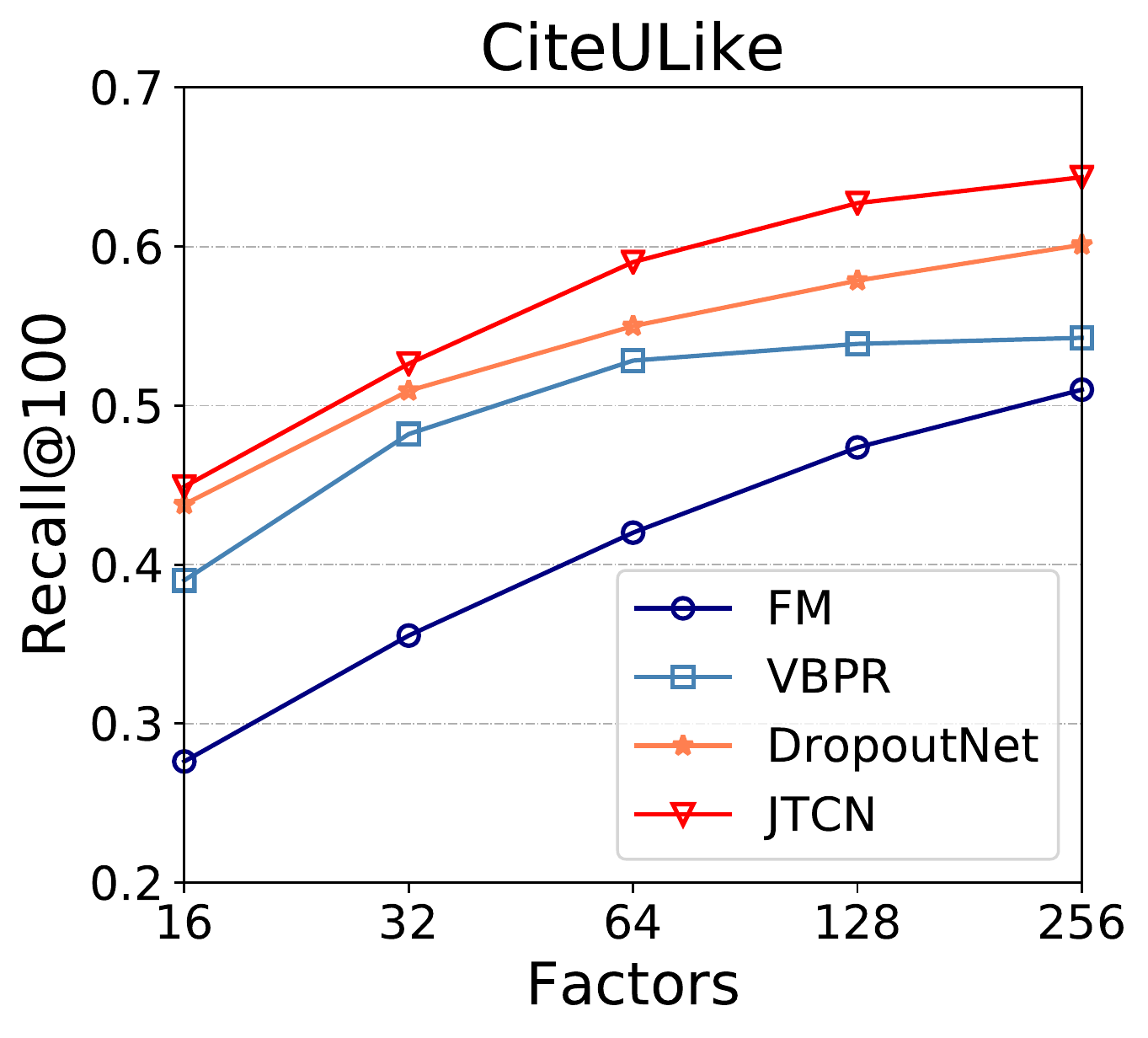}}
\subfigure[CiteULike--NDCG@100]{\includegraphics[width=4.25cm]{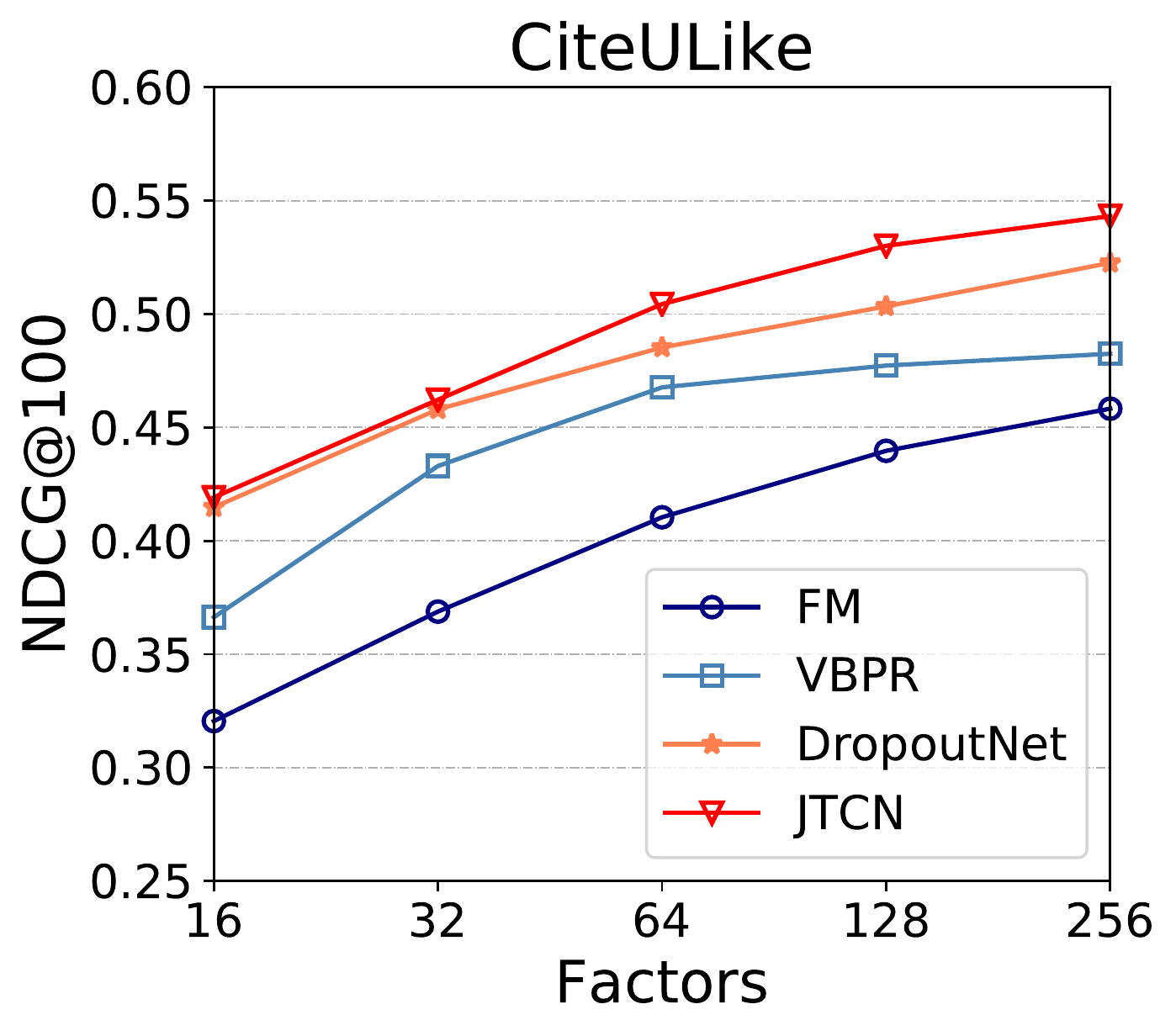}}
\subfigure[Amazon--Recall@100]{\includegraphics[width=4.25cm]{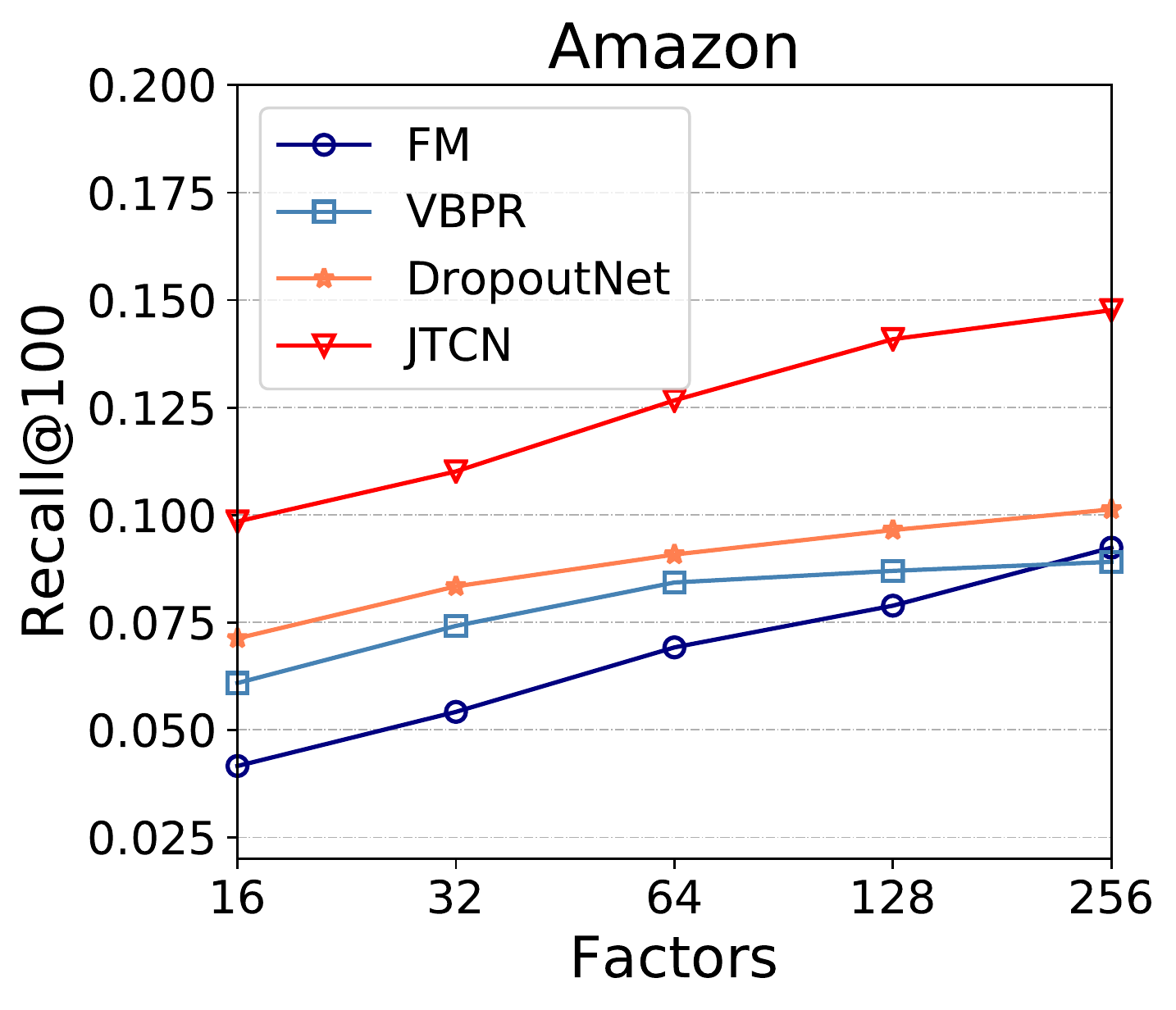}}
\subfigure[Amazon--NDCG@100]{\includegraphics[width=4.15cm]{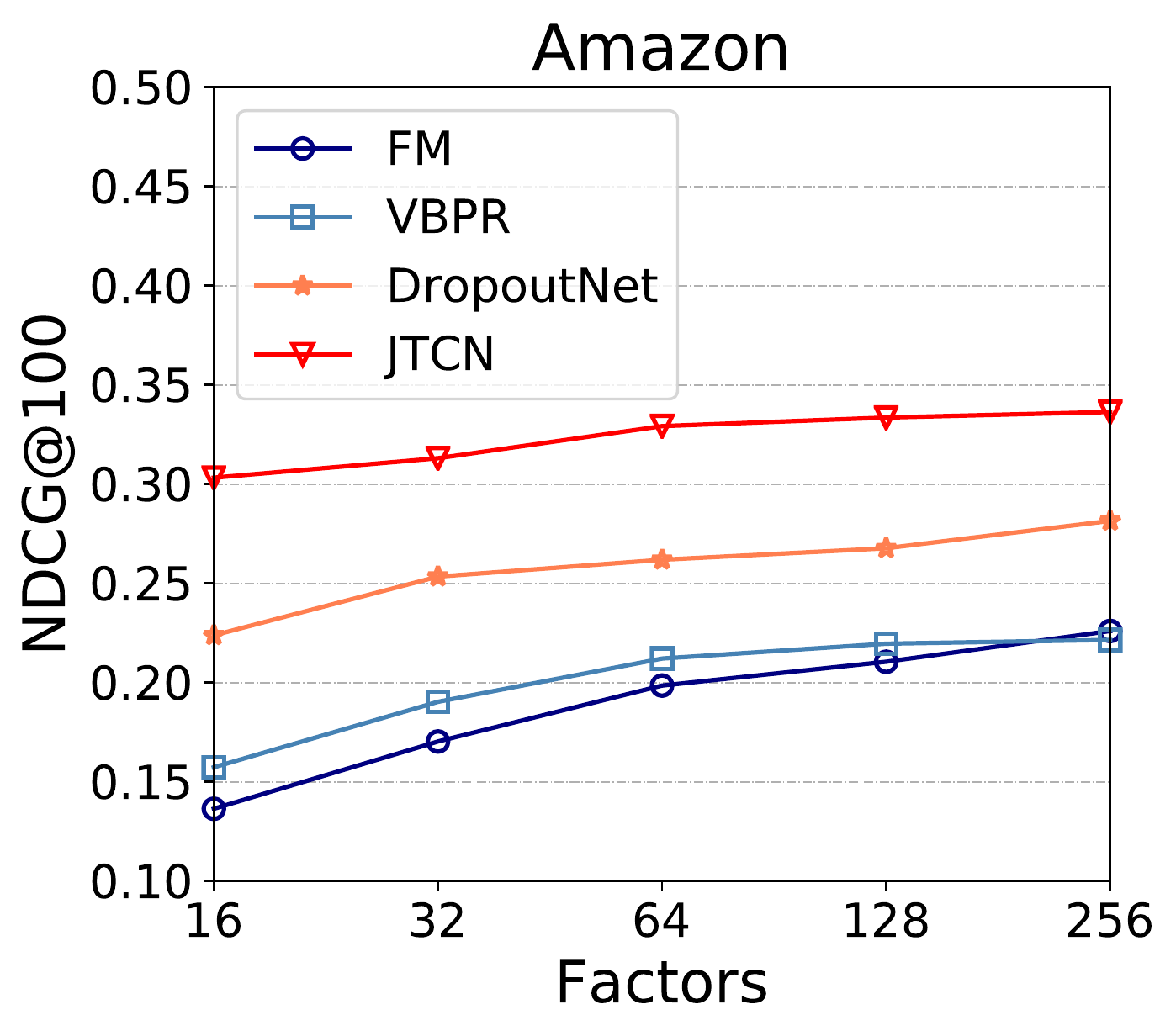}}
% \subfigure[Dynamic Graph]{\includegraphics[width=3.8cm]{fig/dynamic_fig/item_ocean_2_50_circle.pdf}}
% \subfigure[Dynamic Graph]{\includegraphics[width=3.8cm]{fig/dynamic_fig/item_ocean_11_50.pdf}}
\vskip -0.15in
\caption{Performance of Recall@100 and NDCG@100 \emph{w.r.t} the number of predictive factors on the two datasets.}
\vspace{-0.1in}
\label{fig:factor}
\vskip -0.05in
\end{figure}

To analyze the importance of latent factors, we compare the performance of FM, VBPR, and DropoutNet with the proposed {\modelname} with respect to the number of latent factors. As Figure \ref{fig:factor} shows,
{\modelname} consistently outperforms the baselines.
With the increase of the number of latent factors, the performance improvement compared with the best baseline method generally increases. It may be because the combination of content and preference representations is more informative, which requires a relatively larger hidden dimension to incorporate.

%% file: 4con.tex
\vspace{-0.05in}
\section{Conclusion}
%\vspace{-0.05in}
%1. Learning Content-to-Preference Mapping for Cold Start Recommendation
%2. Preference estimator with dynamic routing for cold start recommendation 
%3. Preference estimator with joint loss for cold start recommendation
%4. Joint learning of preference and content for cold start recommendation
In this paper, a novel neural network model, namely joint training capsule network ({\modelname}) is first introduced to harness the advantages of capsule model for extracting high-level user preference in the cold start recommendation task. {\modelname} optimizes the mimic loss and softmax loss together in an end-to-end manner: the mimic loss is used to mimic the preference for cold start users or items; the softmax loss is trained for recommendation. An attentive capsule layer is proposed to aggregate high-level preference from the low-level interaction history via a dynamic routing-by-agreement mechanism. Experiments on two real-world datasets show that our JTCN consistently outperforms baselines.
% for the cold start recommendation task.